\documentclass[10pt]{iopart}
\usepackage{graphicx}
\usepackage{float}
\usepackage{multirow}
\usepackage[breaklinks=true,colorlinks,citecolor=blue,linkcolor=blue,urlcolor=blue]{hyperref}
\begin{document}
\title{Highly accurate wavefunctions for two-electron systems using two parameteres}
\author{Rabeet Singh Chauhan and Manoj K. Harbola}
\address{Department of Physics, Indian Institute of Technology-Kanpur,Kanpur-208 016, India}
\date{\today}
\begin{abstract}
It is shown for two electron atoms that ground-state wavefunctions of the form 
\begin{equation*}
\fl~~~~~~~~~~~\Psi(\vec{r_{1}}, \vec{r_{2}})=\phi(\vec{r_{1}})\phi(\vec{r_{2}})(\cosh ar_{1}+\cosh ar_{2})(1+0.5 r_{12}e^{-b r_{12}})
\end{equation*}
where $\vec{r_{1}}$ and $\vec{r_{2}}$ are the coordinates of two electrons and $r_{12}=|\vec{r_{1}}-\vec{r_{2}}|$, can be made highly accurate by optimizing $a$, $b$ and $\phi$.  This is done by solving a variationally derived equation for $\phi$ for a given $a$ and $b$ and  finding $a$ and $b$ so that the expectation value of the Hamiltonian is minimum.  For the set $\{a, b, \phi\}$ the values for various quantities obtained from the above wavefunction are compared with those given by $204$-parameter wavefunction of Koga \etal\cite{Koga}and are found to be matching quite accurately(within ppm) with them.    
\end{abstract}
\section{Introduction}
The simplest of many-electron systems are those with two electrons like  ions of the helium isoelectronicc series or a harmonium atom.  Since the Schr$\ddot{o}$dinger equation for these systems cannot be solved analytically, accurate solutions are obtained by using the variational principle for the energy.  Thus an approximate parametrized form of the wavefunction is chosen incorporating the effect of electron-electron interactionon in the wavefunction and the expectation value of the Hamiltonian (atomic units are used)
\begin{equation}
H = -\frac{1}{2}\nabla_{1}^{2}-\frac{1}{2}\nabla_{2}^{2}+v_{ext}(\vec{r_{1}})+v_{ext}(\vec{r_{2}})+\frac{1}{r_{12}}
\end{equation}
is minimized with respect to the parameters in the wavefunction. This leads to an optimized approximate wavefunction and the corresponding energy.  The various forms used for the wavefunctions are those by Hylleraas\cite{HY,HY1,HY2,HY3,HY4}, Chandrasekhar\cite{Chandra} and  Kinoshita\cite{Kino1,Kino2}.  As an example of how these forms incorporate the effect of electron-electron interaction, we consider the Chandrasekhar wavefunction for two-electron atom given as 
\begin{equation}
\Psi_{C}(\vec{r_{1}}, \vec{r_{2}})=C_{N} (e^{-a r_{1}}e^{-b r_{2}}+e^{-a r_{2}}e^{-b r_{1}})
\end{equation}
where $a$ and $b$ are the parameters to be optimized and $C_{N}$ is the normalization constant.  The idea behind writing the wavefunction in the form above is as follows.  In the ground state of a two electron atom, each electron moves essentially in a hydrogen-atom like $1s$ orbital (of the form $e^{-ar}$)but because of electron closer to nucleus, the effective nuclear charge is different for the outer electron.  Hence the parameters for the orbitals of the two electrons are different.  Symmetrization of the resulting product wavefunction then leads to the form given in equation $(2)$.
Another way of obtaining accurate wavefunctions has been given by Hylleraas\cite{HY,HY1,HY2,HY3,HY4} where the wavefunction is expressed as 
\begin{equation}
\Psi_{H}(\vec{r_{1}}, \vec{r_{2}})=e^{-\zeta s}\sum_{i=1}^{N} c_{i} s^{l_{i}} t^{m_{i}} u^{n_{i}}
\end{equation}
where $s=r_1+r_2$,~~~$t=r_1-r_2$,~~~~ $u=|\vec{r_{1}}-\vec{r_{2}}|$, and  $\zeta$ and $c_{i}$ are variational parameters.  The most accurate calculation using this form of the wavefunction has been that of Frankowski and Pekeris\cite{Frank}, who used a $246$-term wavefunction to get  highly accurate energies for the helium isoelctronic series.  Employing a $230$-term wavefunction Freund et al. \cite{Frue} got essentially the same results as those of \cite{Frank}.  Using another modification to Hylleraas type wavefunction, Koga \etal \cite{Koga} constructed a $204$-term wavefunction that gives energy within $16$ nano-Hartree of the energy given in \cite{Frue}.  In addition, usefulness of the work of \cite{Koga} lies in the fact that they expressed the electronic density of the $204$-parameter wavefunction in a simple analytical form with $43$ parameters.  Along similar lines, in $1994$  Umrigar and Gonze\cite{Cyru} constructed very accurate densities for He isoelectronic series by using $491$-term wave function having the same form of the basis set as used by Freund \etal.  The resulting densities have been used in the past to construct exchange-correlation potential\cite{Cyru} and to calculate\cite{mkh1} polarizability of two electron atoms and also to study derivative discontinuity of the exchange-correlation\cite{mkh2} of potential of density functional theory.  Examples of references \cite{Cyru,mkh1,mkh2} are sufficient to show that the availability of accurate densities facilitate studies in fundamental aspects of density functional theory\cite{Yang}.  As such methods that can give accurate densities with relative ease are desirable to facilitates such studies.  With this in mind Le Sech\cite{Lese} introduced a semianalytical wavefunction for the ground state of two electron atoms and ions that gives energy within parts per million (ppm) of the answers of references \cite{Frank,Frue,Koga}. 
\section{Le Sech wavefunction}
 The wavefunction proposed by Le Sech is motivated by the intuitive aspects of the Chandrashekhar wavefunction and correlation factor given by earlier studies \cite{Hirsch}.  The wavefunction is given as
\begin{equation}
\Psi_{L}(\vec{r_{1}}, \vec{r_{2}})=C_{N}e^{-Z(r_1+r_2)}(\cosh ar_{1}+\cosh ar_{2})(1+0.5 r_{12}e^{-b r_{12}})  \label{psil}
\end {equation}
where $C_{N}$ is the normalization constant and $Z$ is the atomic number of the atomic system.  The factor $(\cosh ar_{1}+\cosh ar_{2})$ is chosen so that cusp condition\cite{Kato} is satisfied exactly by the wavefunction.  The parameters $a$ and $b$ are obtained variationally by minimizing the expectation value of the Hamiltonian with respect to them.  The energies so obtained for H\textsuperscript{-}, He, Li\textsuperscript{+}, Be\textsuperscript{2+}, B\textsuperscript{3+} are given in \Tref{tb1} along with the exact energies obtained in \cite{Koga}.  The differences between the two energies are also shown. We mention that the values of parameters $a$ and $b$ and the corresponding energies obtained by us are slightly different from those reported in ref \cite{Lese}.  It is clear from \Tref{tb1} that the wavefunction $\Psi_{L}$ of \Eref{psil} leads to accurate energies with the difference from those of \cite{Koga} becoming smaller with increasing Z.  The difference for H\textsuperscript{-} is $1899$ parts per million  and goes down to $104(ppm)$for $B\textsuperscript{3+}$.  We note that the wavefunction of \Eref{psil} can be improved further with the inclusion of one more parameter and leads to a closer agreement with the exact energies. \\ 
\begin{table}[h] 
\caption{\label{tb1} Energies for He-like systems using approximate wavefunction $\Psi_L$ of \Eref{psil} and $\Psi_{ML}$ of \Eref{psiml}.  Their comparision with the energies given in \cite{Koga}(given in the column under$\Psi_K$) is also made in the table.  The differences of each energy from that given in \cite{Koga} is given in ppm under the energies.}
\begin{center}
\resizebox{0.9\textwidth}{!}{
\begin{tabular}{|l|l|l|c|c|l|l|c|c|}
\hline
Atom &  \multicolumn{3}{c|}{$\Psi_{L}$}& $\Psi_K$ &\multicolumn{3}{c|}{$\Psi_{ML}$}     \\
\cline{2-8}
                      &   a  &   b    & -Energy(a.u.)& -Energy(a.u.)& -Energy(a.u.) &   a   &  b        \\   \hline        
 H\textsuperscript{-} & 0.58 & 0.06   &   0.5267   &    0.5277  &    0.5271   & 0.62  & 0.06      \\
                      &      &        &    (1899)  &            &    (1137)   &       &           \\   
 He                   & 0.72 & 0.20   &   2.9020   &    2.9037  &    2.9028   & 0.93  & 0.20      \\
                      &      &        &    (586)   &            &    (310)    &       &           \\  
 Li\textsuperscript{+}& 0.87 & 0.36   &   7.2778   &    7.2799  &    7.2788   & 1.19  & 0.36      \\
                      &      &        &    (289)   &            &    (151)    &       &           \\  
Be\textsuperscript{2+}& 0.99 & 0.52   &   13.6533  &   13.6555  &   13.6544   & 1.48  & 0.54      \\
                      &      &        &    (161)   &            &     (81)    &       &           \\  
 B\textsuperscript{3+}& 1.10 & 0.67   &   22.0286  &   22.0309  &   22.0297   & 1.72  & 0.70      \\
                      &      &        &    (104)   &            &     (55)    &       &           \\   
\hline                        
\end{tabular}}
\end{center}
\end{table}

The question we now ask is how accurate are the densities given by $\Psi_{L}$?  This is important from the point of view of having an accurate  and easily accesible density of two electron systems if we were to employ them to investigate the foundational aspects of density functional theory of many electron systems in general and two electron systems in particular.  To this end we present in \Tref{tb2} the values of $\rho(r=0)$, $\frac{d\rho}{dr}|_{r=0}$ and in Table \ref{tb3} and \ref{tb4} various moments of the densities obtained from the Le Sech wavefunction.  A comparision of the results is made  with the accurate values of those obtained from the densities given in \cite{Koga}.  It is seen that the densities given by $\Psi_{L}$ are smaller than the exact values of the densities at $r=0$.  Since the Kato-cusp condition is satisfied by both $\Psi_{L}$ and $\Psi_{K}$, the derivative $|\frac{d\rho}{dr}|_{r=0}$ are also small in comparision with the exact density.  This implies that the Le Sech densities  may be little more spread out than the exact densities.  
\begin{table}[h]
\caption{\label{tb2}Densities and its derivative at $r=0$ are calculated by using $\Psi_{L}$, $\Psi_{ML}$. For the density given by Koga \etal(given in the column under $\psi_K$) we have used the exact expression given in their paper\cite{Koga} to calculate $\rho(r=0)$ and $-\frac{d\rho}{dr}(r=0)$. We see that Le Sech density and its derivatives are  smaller than the density given in\cite{Koga} at $r=0$ but the Modified Le Sech density is close to them.  Numbers given are in atomic units.}
\begin{center}
\resizebox{0.8\textwidth}{!}{
\begin{tabular}{|l|l|l|c|c|l|l|c|c|}
\hline
Atom &  \multicolumn{3}{c|}{$\rho(r=0)$} &\multicolumn{3}{c|}{$-\frac{d\rho}{dr}(r=0)$}\\
\cline{2-7}
     &  $\Psi_L$  &   $\Psi_{K}$  &  $\Psi_{ML}$  & $\Psi_L$ &  $\Psi_{K}$   & $\Psi_{ML}$      \\ \hline         
H\textsuperscript{-}   & 0.310  & 0.323  & 0.329   & 0.621   & 0.649   & 0.657      \\
He                     & 3.554  & 3.621  & 3.616   & 14.214  & 14.483  & 14.462      \\
Li\textsuperscript{+}  & 13.552 & 13.704 & 13.693  & 81.312  & 82.223  & 82.185      \\
Be\textsuperscript{2+} & 34.180 & 34.396 & 34.384  & 273.438 & 275.167 & 275.077      \\
B\textsuperscript{3+}  & 69.164 & 69.517 & 69.501  & 691.640 & 695.165 & 695.011      \\
\hline
\end{tabular}}
\end{center}
\end{table}
\begin{table}[h]  
~~~~~ Next in \Tref{tb3} and \Tref{tb4}, we present various moments of the density obtained from $\Psi_L$ of \Eref{psil} and compare them with their exact counterparts.  It is seen that $\langle r^{-2} \rangle $ and $\langle r^{-1} \rangle $ for $\Psi_L$ are less than those obtained from\cite{Koga} the exact density.  This is consistent with the density for $\Psi_L$ being smaller than the exact one near the nucleus.  For $\langle r \rangle $, $\langle r^{2} \rangle $ and $\langle r^{3} \rangle $ on the hand, it is observed that their values are larger than the exact moments for most of the cases (exception is for H\textsuperscript{-}). This again shows that for $\Psi_L$, the density is more spread out in general.
\caption{\label{tb3}Comparision of moments calculated by using $\Psi_{L}$ of \Eref{psil} and $\Psi_{ML}$ of \Eref{psiml} with the moments given in \cite{Koga} for two electron atoms.  The latter are given in column under $\Psi_K$.  Numbers are given in atomic units.}
\begin{center}
\resizebox{0.8\textwidth}{!}{
\begin{tabular}{|c|c|c|c|c|c|c|}
\hline
Atom & \multicolumn{3}{c|}{$\langle r^{-2} \rangle $}&\multicolumn{3}{c|}{$\langle r^{-1} \rangle $}\\
\cline{2-7}
 &   $\Psi_L$ &  $\Psi_{K}$   & $\Psi_{ML}$  & $\Psi_L$ &  $\Psi_{K}$   & $\Psi_{ML}$      \\ \hline         
H\textsuperscript{-}   & 2.145  & 2.233  & 2.229   & 1.348  & 1.366 & 1.367    \\
He                     & 11.878 & 12.035 & 12.002  & 3.363  & 3.377 & 3.376    \\
Li\textsuperscript{+}  & 29.574 & 29.855 & 29.762  & 5.362  & 5.376 & 5.375    \\
Be\textsuperscript{2+} & 55.301 & 55.680 & 55.485  & 7.367  & 7.376 & 7.376    \\
B\textsuperscript{3+}  & 88.900 & 89.507 & 89.146  & 9.366  & 9.375 & 9.375    \\
\hline
\end{tabular}}
\end{center}
\end{table}
\begin{table}[h]
\caption{\label{tb4}The caption is same as that of \Tref{tb3}}
\begin{center}
\resizebox{0.9\textwidth}{!}{
\begin{tabular}{|c|c|c|c|c|c|c|c|c|c|}
\hline
Atom &  \multicolumn{3}{c|}{$\langle r\rangle $} &  \multicolumn{3}{c|}{$\langle r^{2} \rangle $} &\multicolumn{3}{c|}{$\langle r^{3} \rangle $}\\
\cline{2-10}
 &  $\Psi_L$  &   $\Psi_{K}$ & $\Psi_{ML}$  & $\Psi_L$  &   $\Psi_{K}$  &  $\Psi_{ML}$  & $\Psi_L$ &  $\Psi_{K}$   & $\Psi_{ML}$       \\ \hline         
H\textsuperscript{-}   & 5.331 &5.420 & 5.422 & 22.000& 23.830& 23.859 & 125.030& 152.000 & 152.366    \\
He                     & 1.863 &1.859 & 1.859 & 2.402 & 2.387 & 2.386  & 3.996  & 3.936   & 3.934    \\
Li\textsuperscript{+}  & 1.148 &1.146 & 1.145 & 0.898 & 0.893 & 0.892  & 0.895  & 0.883   & 0.882    \\
Be\textsuperscript{2+} & 0.829 &0.829 & 0.829 & 0.465 & 0.464 & 0.464  & 0.331  & 0.328   & 0.328    \\
B\textsuperscript{3+}  & 0.650 &0.649 & 0.649 & 0.285 & 0.284 & 0.284  & 0.157  & 0.156   & 0.157    \\
\hline
\end{tabular}}
\end{center}
\end{table}
We also test the density in terms of it satisfying the ionization potential ($IP$) theorem  in Kohn-Sham density functional theory.  Furthermore, we also study the accuracy of the corresponding Kohn-Sham exchange-correlation potential.  According to the $IP$ theorem \cite{PPLB,LPS} the highest occupied orbital eigenenergy is equal to negative of the ionization potential of a many electron system.  Accordingly for a Kohn-Sham system for a given ground-state density of two electron system, its eigenvalue should be equal to $I=-(E+\frac{Z^2}{2})$ where $E$ is the energy of the two-electron system.  To test the accuracy of the Le Sech densities, we construct the corresponding Kohn-Sham system using the Zhao-Parr\cite{Zhao} method, which is described briefly in the following. \\ \par
  For the orbitals of noninteracting system of electrons the kinetic energy $T_s$ and ground state density $\rho(r)$ are given by \\
\begin{equation} 
 T_{s}=\sum_{i}\langle\phi_{i}(\mathbf{r})|-\frac{1}{2}\nabla^{2}|\phi_{i}(\mathbf{r})\rangle, ~~~~  \rho(r)=\sum_{i}|\phi_i(r)|^{2} \label{tsrho}
\end {equation}
For a given ground state density, the Kohn-Sham system is formed by minimizing $T_s+\int\mathrm{v_{ext}(r)\rho(r)}\mathrm{d}\mathbf{r}$ with respect to $\phi_i$'s with the constraint. 
\begin{equation}
\int\int\mathrm{\frac{[\rho(r)-\rho_{0}(r)][\rho(r')-\rho_{0}(r')]}{|\mathbf{r}-\mathbf{r'}|}}\mathrm{d}\mathbf{r}\mathrm{d}\mathbf{r'}=0 \nonumber
\end{equation}
which is equivalent to $\rho(r)-\rho_{0}(r)=0$,  and $\int \mathrm{|\phi_i|^2}\mathrm{d}\mathbf{r}=1$.  This leads to the equation  
\begin{equation}
-\frac{1}{2}\nabla^{2}\phi_{i}(\mathbf{r})+v_{ext}(r)\phi_{i}(\mathbf{r})+v(r)\phi_{i}(\mathbf{r})=\epsilon_{i}\phi_{i}(\mathbf{r}) \label{zhaopa}\nonumber
\end{equation}
where
\begin{equation}
v(r)=\lambda\int\mathrm{\frac{[\rho(r')-\rho_{0}(r')]}{|\mathbf{r}-\mathbf{r'}|}}\mathrm{d}\mathbf{r'}  \label{pzp}\nonumber
\end{equation}
where $\lambda$ and $\epsilon_{i}$ are the Lagrangian multipliers to enforce the two constraints above.  \Eref{zhaopa} is solved self-consistently using a Herman-Skillman\cite{Herm} program modified suitably.  Since at the solution point $\rho(r)=\rho_0(r)$, the value of $\lambda$ should go to infinity such that  $v(r)$ remains finite.  In this paper we have performed all the calculations for $\lambda=30000$.  We point out that for two electron systems, the Kohn-Sham systems can also be found directly from the Laplacian of the density.  We however, chose Zhao-Parr method because it is more general and more accurate.

\begin{table}[h]
\caption{\label{tb5}Highest occupied orbital eigenvalue $\epsilon_{max}$ and ionization potential $I=-(E+\frac{Z^{2}}{2})$ for two electron systems.  Numbers given are in atomic units}
\begin{center}
\resizebox{0.7\textwidth}{!}{
\begin{tabular}{|c|c|c|c|c|c|}
\hline
Atom &  \multicolumn{2}{c|}{$ \Psi_L $} &\multicolumn{2}{c|}{$ \Psi_{ML} $}   &  $ I_{expt}\cite{Lide} $ \\
\cline{2-6}
                       &$-\epsilon_{max}$&  $I$     & $-\epsilon_{max}$  &  $I$ &        \\ \hline         
H\textsuperscript{-}   & 0.0613 & 0.0267 & 0.0276   & 0.0271  & 0.0277    \\
He                     & 0.8265 & 0.9020 & 0.9059   & 0.9027  & 0.9036    \\
Li\textsuperscript{+}  & 2.5718 & 2.7778 & 2.7821   & 2.7787  & 2.7798    \\
Be\textsuperscript{2+} & 5.3581 & 5.6533 & 5.6593   & 5.6543  & 5.6557    \\
B\textsuperscript{3+}  & 9.1406 & 9.5286 & 9.5349   & 9.5296  & 9.5320    \\
\hline
\end{tabular}}
\end{center}
\end{table}

Shown in \Tref{tb5} are the values of $-\epsilon_{max}$ and $I$ for $\Psi_{L}$  for the ions studied in this paper.  It is seen for $\Psi_L$ that  the values of $\epsilon_{max}$ and $I$ are comparable to each other.  However, they are not close enough to satisfy the $IP$ theorem.  Furthermore, the values of $I$ are close to the experimental ionization potential, which shows that the total energies $E$ for $\Psi_L$ is accurate.  On the other hand $-\epsilon_{max}$ is not accurate for $\Psi_L$. This is particularily  evident for the densities of the $H^{-}$ ion and He atom where the difference between the two is $121$ and $8$, percent respectively.  Since the eigenvalue $\epsilon_{max}$ is determined by the asymptotic decay of the electron density, it is evident that asymptotically, the density given by Le Sech wavefunction is not accurate.  This is further evidenced by the corresponding Kohn-Sham exchange-correlation shown in figure $1$ where we have plotted the exchange-correlation potential of He for both Le Sech density as well as the density given in\cite{Koga} and another accurate density of Umrigar and Gonze\cite{Cyru}.  It is clear that Le Sech density leads to smaller magnitude of the exchange-correlation potential near the nucleus, which in turn gives densities which are more spread out. \\   \par 
\begin{figure}[h]
\begin{center}
\caption{\label{excc}Comparision of the exchange-correlation potential calculated for given Le Sech density ($Vxc_L$), modified Le Sech density ($Vxc_{ML}$) and for the density given in\cite{Koga} ($Vxc_K$) and for the accurate density of Umrigar and Gonze\cite{Cyru}($Vxc_{UG}$).}
\includegraphics[width=0.9\textwidth]{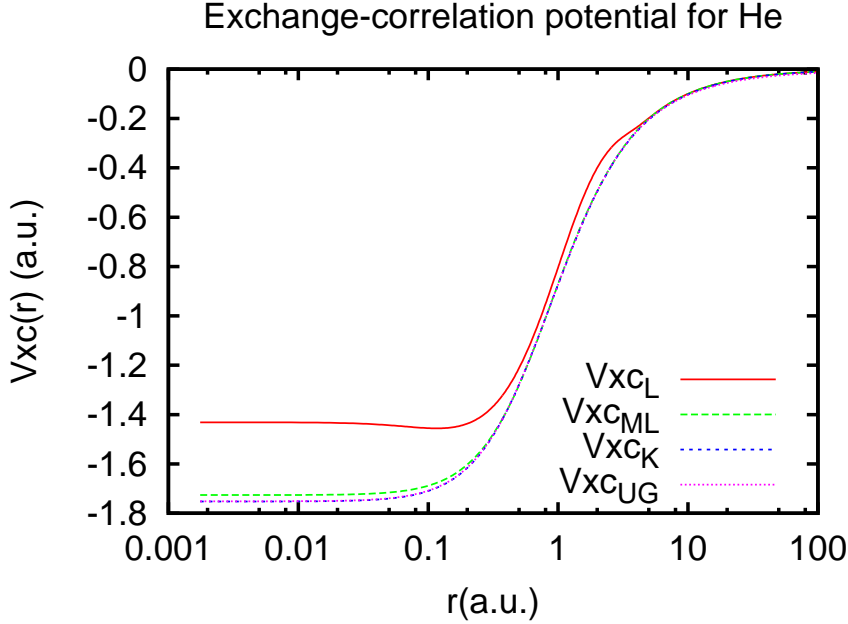} \label{figure}
\end{center}
\end{figure} 
It is evident from the results presented so for that the Le Sech wavefunction needs to be improved if it were to give accurate densities.  The rest of the paper is then devoted to modifying the Le Sech wavefunction so that it leads to densities that can be employed to study exact properties in density functional theory.
\section{Modified Le Sech wavefunction}
In this section we modify the Le Sech wavefunction so that the new wavefunction gives accurate energies as well as densities.  We first point out how the Le Sech wavefunction can be made better.  We feel that keeping the parameter $Z$ (of the $e^{-Z r}$ factor) fixed in the Le Sech wavefunction does not provide enough flexibility in the wavefunction and it is this component of the wavefunction that needs to be calculated more accurately.  Hence the modified Le Sech (ML) wavefunction for the ground state of two electron atoms that we propose is  
\begin{equation}
\Psi_{ML}(\vec{r_{1}}, \vec{r_{2}})=\phi(r_1)\phi(r_2)(\cosh ar_{1}+\cosh ar_{2})(1+0.5 r_{12}e^{-b r_{12}})  \label{psiml}
\end{equation}
where for a given set of $a$ and $b$, the orbital $\phi(r)$ is to be calculated self-consistently by solving the equation derived below which is obtained from the Euler equation 
\begin{equation}
\frac{\delta}{\delta \phi(r)}[\langle H \rangle_{ML}-E\langle \Psi_{ML}|\Psi_{ML}\rangle ]=0   \label{euler}
\end{equation}
where $\langle H \rangle_{ML}$ is the expectation value of the Hamiltonian of equation (2) with respect to the wavefunction $\Psi_{ML}$ above. \Eref{euler} leads to following Schr$\ddot{o}$dinger-like equation for $\phi(r)$ (in the equation below $r_{12}=|\vec{\mathbf{r}}-\vec{\mathbf{r_2}}|$, $\nabla$ is gradient operater in terms of $\vec{r}$ and $\nabla_2$ in terms of $\vec{r_2}$).
\begin{eqnarray} \label{maineq}\nonumber
\fl-\frac{1}{2}\nabla^{2}\phi(\mathbf{r})-\frac{1}{A(r)}\nabla\phi(\mathbf{r})\cdot\int\mathrm{f(r, r_{2}, r_{12})|\phi(r_{2})|^{2}\nabla f(r, r_{2}, r_{12})}\mathrm{d}\mathbf{r_{2}}  \\   \nonumber
\fl~~~~~-\frac{1}{2 A(r)}\int[\mathrm{|\phi(r_{2})|^{2}f(r, r_{2}, r_{12})\nabla^{2} f(r, r_{2}, r_{12})}+\mathrm{|\phi(r_{2})|^{2}f(r, r_{2}, r_{12})\nabla_{2}^{2} f(r, r_{2}, r_{12})}  \\         \nonumber
\fl~~~~~~~~+\mathrm {|f(r, r_{2}, r_{12})|^{2}\phi(r_{2})\nabla_{2}^{2}\phi(r_{2})}+\mathrm{2 f(r, r_{2}, r_{12})\phi(r_{2})\nabla_{2}\phi(r_{2})\cdot\nabla_{2}f(r, r_{2}, r_{12})}]\mathrm{d}\mathbf{r_{2}}\phi(\mathbf{r})\\   \nonumber
-\frac{Z}{r}\phi(\mathbf{r})-\frac{Z}{A(r)}\int\mathrm{\frac{|\phi(r_{2})f(r, r_{2}, r_{12})|^{2}}{r_{2}}}\mathrm{d}\mathbf{r_{2}}\phi(\mathbf{r})\\ 
~~~~~~~~~~~~+\frac{1}{A(r)}\int\mathrm{\frac{|\phi(r_{2})f(r, r_{2}, r_{12})|^{2}}{|\mathbf{r}-\mathbf{r_{2}}|}}\mathrm{d}\mathbf{r_{2}}\phi(\mathbf{r})=E \phi(\mathbf{r})   
\end{eqnarray}
where
\begin{equation}
 f(r, r_2, r_{12})=(\cosh ar+\cosh ar_{2})(1+0.5 r_{12}e^{-b r_{12}}) \nonumber
\end{equation}
and 
\begin{equation}
 A(r)=\int\mathrm{|\phi(r_{2})f(r, r_{2}, r_{12})|^{2}}\mathrm{d}\mathbf{r_{2}}    \nonumber
\end{equation}
The equation above is solved self-consistently for a range of parameters a and b.  The solution gives $\phi(r)$ and the corresponding energy E as the eigenvalue.  The best wavefuncion corresponds to the values of a and b that give the minimum value of E.  \\ \par
   We wish to point out that after completing our work, we discovered that the above equation was solved\cite{Babe} in $1937$ to obtain ground state energy and polarizability of the He atom. However the scope of the present paper is much wider.  In this paper we investigate the wavefunction not only for the energy but also for the density it gives rise to.  We present the results obtained by solving \Eref{maineq} for the  two electron atomic systems studied so for and show the accuracy of the energy as well as the density that is achieved by self-consistently determined $\Psi_{ML}$.  From the results presented below, it becomes clear that \Eref{maineq} provides a computationally straightforward method of obtaining highly accurate wavefunctions for two-electron systems.  The method can also be applied with equal ease to other two-electrons systems such as electrons in a confined environment or harmonic oscillator potential. \\ \par
  Given in \Tref{tb1} are the results for the total energies of the modified Le Sech wavefunction $\Psi_{ML}$ of \Eref{psiml}.  As is evident, for the modified wavefunction, the total energies are better than those for the Le Sech functions.  The modified wavefuction reduces the error of energies from Le Sech wavefunction by roughly a factor of $2$.  This is because the wavefunctions has been made more flexible.  In addition to giving improved energies, $\Psi_{ML}$ also gives much better densities than $\Psi_L$. We present these results in the following. \\ \par
In \Tref{tb2}, we also present the density and its derivative at $r=0$ obtained from $\Psi_{ML}$ of \Eref{psiml}.   We see that the values of modified Le Sech density and its derivative are very close the values obtained using exact expression given in \cite{Koga}.  In  \Tref{tb3} and \Tref{tb4} we present the various moments of the density obtained by $\Psi_{ML}$. These moments are compared with the moments given in \cite{Koga}.  We get all the moments calculated by the modified Le Sech density close to the moments given in \cite{Koga}. This implies that the modified Le Sech density is very close to the exact density. \\ \par 
 In \Tref{tb5}, $IP$ theorem is tested for both Le Sech and modified Le Sech density, where the $\epsilon_{max}$ are calculated by solving Equation (7). We see that the modified Le Sech density satifies the $IP$ theorem almost exactly.  Finallly  in \Fref{figure} we show the exchange-correlation potential  by using  Le Sech density, modified Le Sech density, density given in\cite{Koga} and for the density obtained by Umrigar and Gonze\cite{Cyru}.  We see that the exchange-correlation potential given by modified Le Sech densities is very close to the exact potential.   

\section{Conclusion}
Le Sech wavefunction\cite{Lese}, which gives very accurate energies for He-like systems, has been modified so that both the energies and densities obtained from the modified wavefunction are highly accurate.  The densities so obtained can be used with confidence to perform fundamental density functional theory investigations.  Along this line of investigation we are now employing the method proposed to study adiabatic connection in density functional theory.
\ack
We are grateful to Prof.  Cyrus. J. Umrigar for providing exact densities for two electron atoms. 

\end{document}